\documentclass[twocolumn]{aastex63}

\shorttitle{In the Presence of a Wrecking Ball}
\shortauthors{Stephen R. Kane \& Sarah Blunt}

\begin{document}

\title{In the Presence of a Wrecking Ball: Orbital Stability in the
  HR~5183 System}

\author{Stephen R. Kane}
\affiliation{Department of Earth and Planetary Sciences, University of
  California, Riverside, CA 92521, USA}
\email{skane@ucr.edu}

\author{Sarah Blunt}
\affiliation{Cahill Center for Astronomy \& Astrophysics, California
  Institute of Technology, Pasadena, CA 91106}
\affiliation{Center for Astrophysics, Harvard \& Smithsonian,
  Cambridge, MA 02138, USA}
\affiliation{NSF Graduate Research Fellow}


\begin{abstract}

Discoveries of exoplanets using the radial velocity method are
progressively reaching out to increasingly longer orbital periods as
the duration of surveys continues to climb. The improving sensitivity
to potential Jupiter analogs is revealing a diversity of orbital
architectures that are substantially different from that found in our
solar system. An excellent example of this is the recent discovery of
HR~5183b; a giant planet on a highly eccentric ($e = 0.84$)
$\sim$75~year orbit. The presence of such giant planet orbits are
intrinsically interesting from the perspective of the dynamical
history of planetary systems, and also for examining the implications
of ongoing dynamical stability and habitability of these systems. In
this work, we examine the latter, providing results of dynamical
simulations that explore the stable regions that the eccentric orbit
of the HR~5183 giant planet allows to exist within the Habitable Zone
of the host star. Our results show that, despite the incredible
perturbing influence of the giant planet, there remain a narrow range
of locations within the Habitable Zone where terrestrial planets may
reside in long-term stable orbits. We discuss the effects of the giant
planet on the potential habitability of a stable terrestrial planet,
including the modulation of terrestrial planet eccentricities and the
periodically spectacular view of the giant planet from the terrestrial
planet location.

\end{abstract}



\section{Introduction}
\label{intro}

As both the time baseline and precision of radial velocity (RV)
instruments improve, detections of increasingly longer period planets
are occurring
\citep{wright2008,wittenmyer2011a,wittenmyer2016c,kane2019b}. The
importance of discovering such long-period giant planets is placing of
our solar system architecture in context via measuring occurrence
rates of Jupiter analogs
\citep{boisse2012a,wittenmyer2013a,kipping2016b,buchhave2018}, the
influence of giant planets beyond the snow-line on terrestrial planet
habitability
\citep{raymond2006a,horner2008a,georgakarakos2018,hill2018,sanchez2018},
and possible targets for direct imaging observations
\citep{lannier2017,kane2018c}. A further investigative process that is
invoked in the context of long-period planets is the exploration of
orbital eccentricity distributions \citep{kane2012d,kipping2013b},
which in turn relates to the dynamical histories of systems with
eccentric planets \citep{kane2014b,carrera2016}. The discovery of
particularly long-period giant planets with exceptionally eccentric
orbits thus present opportunities to study critical aspects of
exoplanetary dynamical and habitable evolution.

The most extreme such case detected thus far using the RV method is
that of HR~5183b \citep{blunt2019}. The host star (alias HD~120066) is
a slightly evolved G0 star located at a distance of 31.49~pc. The
planet has one of the longest orbital periods known amongst exoplanets
of $\sim$75~years and orbits the star with an eccentricity of $e =
0.84$. Despite the long orbit, the detection of RV variations during
periastron passage allowed the confirmation of the discovery. Although
no other planets have yet been detected in the system, it is useful to
explore if orbits interior to the giant planet can retain their
long-term dynamical integrity. The long period of the giant planet
combined with with the relatively high eccentricity of the orbit means
that the planet could be aptly described as a ``wrecking ball'', with
its gravitational influence cyclically permeating throughout the
system. A system that consists of a combination of the known planet
and terrestrial planets within the star's Habitable Zone (HZ) would be
a remarkable test-case for the effects of extreme planetary orbits on
the overall architecture and habitability of such systems.

Here we provide a study of the HR~5183 system that includes
calculations of the HZ, dynamical simulations that scan possible
stable orbit locations for additional terrestrial planets, and a
discussion of the implications of extreme orbital architectures.  This
study addresses the question: does the presence of the wrecking
ball planet pose a significant threat to orbital stability of any
potentially habitable planets in the system? In Section~\ref{system}
we describe the system architecture including possible formation
scenarios, and calculate the extent of the system
HZ. Section~\ref{stability} provides the details of an extensive
dynamical simulation that explores regions of stability with the
HZ. These simulations lead to predictions of where terrestrial planets
may possibly reside, described in Section~\ref{predict}, including an
extended discussion of potentially habitable conditions for such
planets under the influence of the known giant planet. We provide
concluding remarks and suggestions for further investigations in
Section~\ref{conclusions}.


\section{System Architecture and Habitable Zone}
\label{system}

\begin{figure*}
  \begin{center}
    \begin{tabular}{cc}
      \includegraphics[angle=270,width=8.5cm]{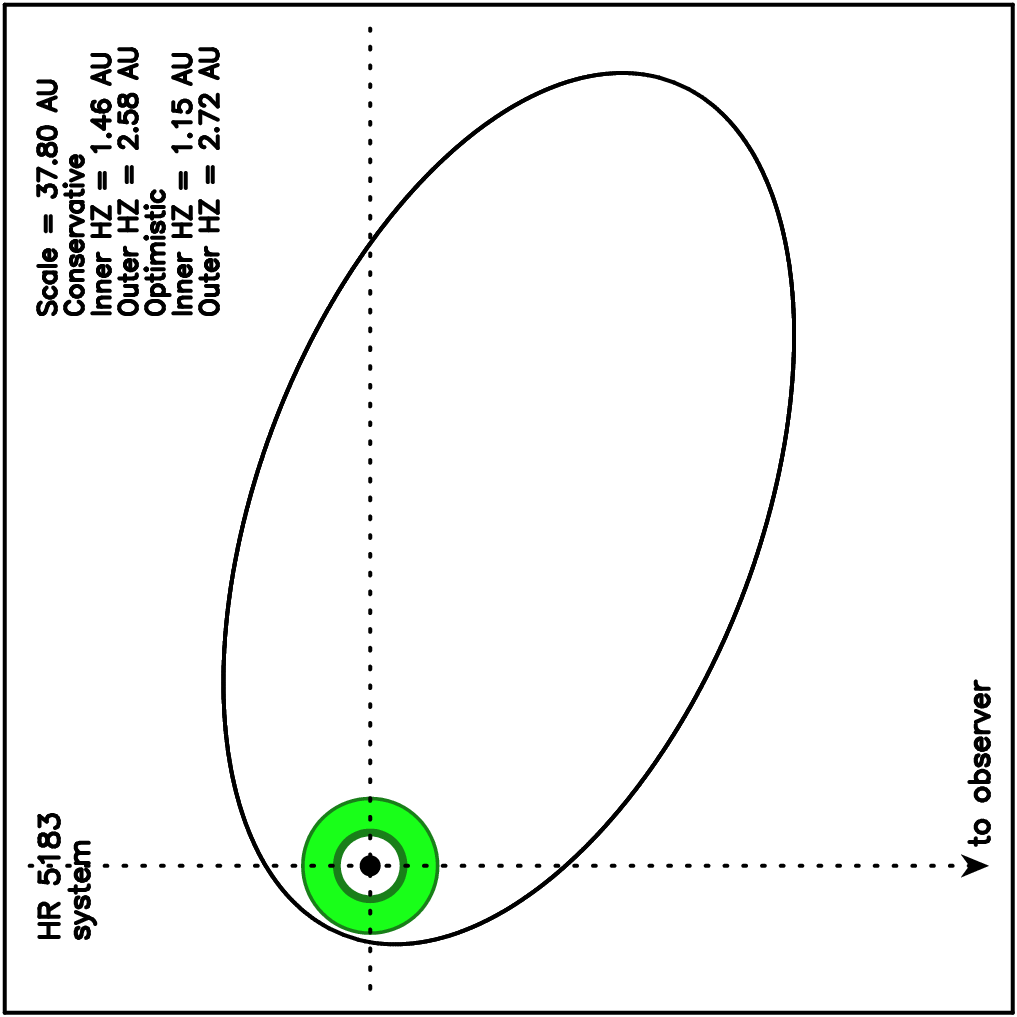} &
      \includegraphics[angle=270,width=8.5cm]{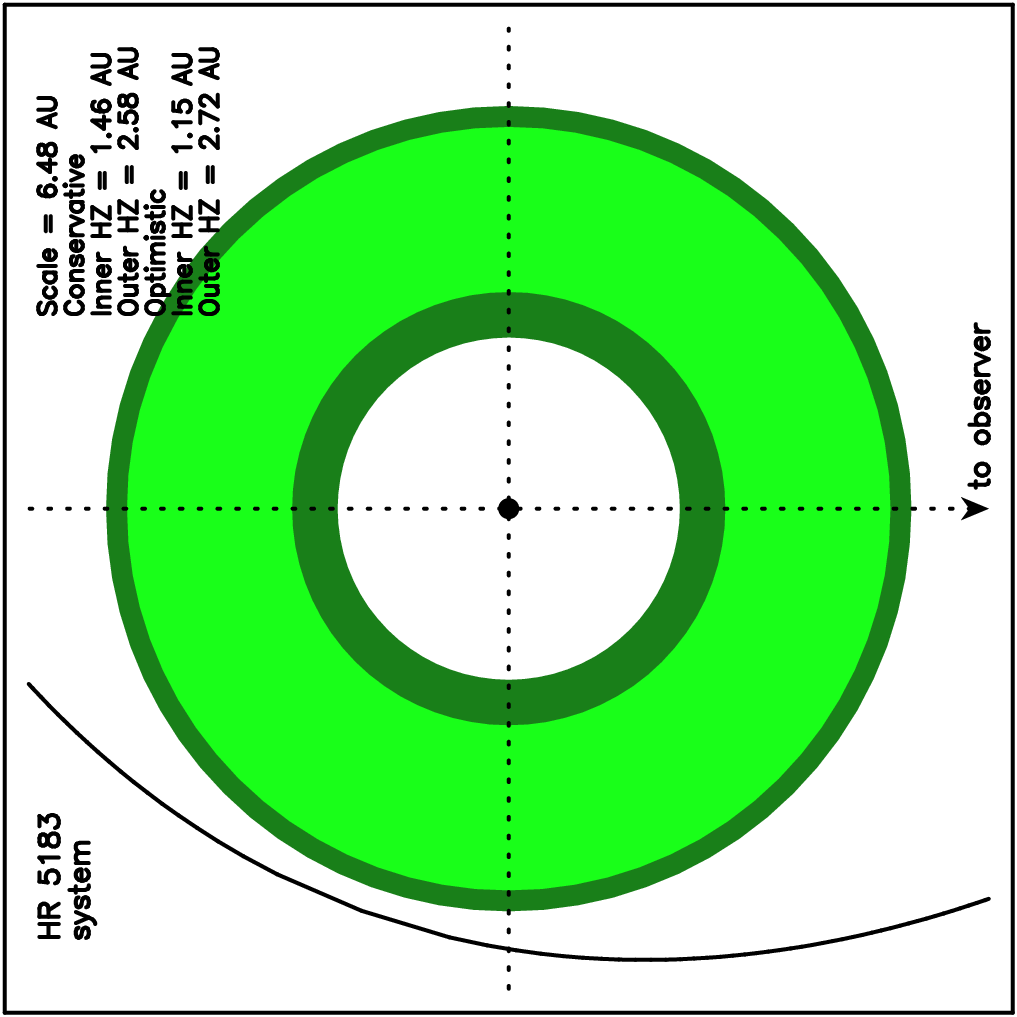}
    \end{tabular}
  \end{center}
  \caption{A top-down view of the HR~5183 system, showing the host
    star (intersection of the dotted cross-hairs) and the orbit of the
    known giant planet (solid line). The extent of the HZ is shown in
    green, where light green is the conservative HZ and dark green is
    the optimistic extension to the HZ. The left panel includes the
    entire orbit of HR~5183b relative to the system HZ. The right
    panel is zoomed in on the HZ and highlights the proximity of the
    periastron of the planet to the outer edge of the HZ.}
  \label{systemfig}
\end{figure*}

The full details of the HR~5183b orbit are provided by
\citet{blunt2019}. The relevant properties for our dynamical analysis
are the stellar mass ($M_\star = 1.07\pm0.04$~$M_\odot$), planet mass
($M_p \sin i = 3.23^{+0.07}_{-0.06}$~$M_J$), semi-major axis ($a =
18\pm2$~AU), orbital eccentricity ($e = 0.84\pm0.02$), and argument of
periastron ($\omega = 339.4\pm0.8\degr$). Note that the planet mass is
a minimum mass depending on orbital inclination. These parameters
result in periastron and apastron distances of 2.89~AU and 33.11~AU
respectively.

The HZ boundaries are derived from Earth-based climate models that
calculate the radiative balance for which surface liquid water is
retained, described in detail by
\citet{kopparapu2013a,kopparapu2014}. The conservative HZ region
extends from an inner boundary, defined by the occurrence of a runaway
greenhouse, to an outer boundary, defined by the location where
maximum CO$_2$ greenhouse occurs \citep{kane2016c}. Similarly, the
optimistic HZ region is an empirical extension to the conservative HZ
region based on assumptions regarding retention of surface liquid
water in the Venusian and Martian evolutionary histories
\citep{kane2016c}. Calculation of the HZ boundaries depend sensitively
on the stellar parameters \citep{kane2014a}, which we adopt from
\citet{blunt2019}. The stellar parameters of $T_\mathrm{eff} = 5794$~K
and $R_\star = 1.53$~$R_\odot$ result in a luminosity of $L_\star =
2.37$~$L_\odot$. Using these stellar parameters, we calculate
conservative HZ boundaries of 1.46~AU and 2.58~AU, and optimistic HZ
boundaries of 1.15~AU and 2.72~AU. Figure~\ref{systemfig} shows a
top-down view of the HR~5183 system, including the orbit of the known
planet and the extent of the HZ. As can be seen in the right panel,
the known giant planet almost brushes against the outer edge of the
HZ.

The presence of such an extreme planetary orbit likely has an
associated turbulent dynamical history. The highest known exoplanet
eccentricity is that of $e = 0.97$ for HD~20782b
\citep{jones2006,kane2016b}. Several studies have suggested that
planet-planet scattering events may be a major contributor to the
observed distribution of eccentric exoplanetary orbits
\citep{chatterjee2008,ford2008c,carrera2019b}. For long-period
eccentric planets, close encounters with other stars (such as wide
binary companions) may serve as the perturbing influence that
contributes the required angular momentum to produce highly eccentric
orbits \citep{kaib2013}. Analysis of ancillary imaging and astrometry
data by \citet{blunt2019} indicate that HR~5183 likely does not have a
stellar companion, or at least not one that would significantly
influence the orbit of HR~5183b. Therefore a planet-planet scattering
scenario is the preferred explanation for the observed eccentricity of
HR~5183b.


\section{Orbital Stability for Terrestrial Planets}
\label{stability}

Our dynamical study of the HR~5183 system seeks to locate stable
orbits within the HZ, such as has previously been determined for the
70~Virginis \citep{kane2015a}, Kepler-68 \citep{kane2015b}, and
HD~47186 \citep{kopparapu2009} systems. As shown in those cases, the
presence of an eccentric giant planet does not exclude terrestrial
planets, and in fact giant planets likely play a contributing role in
the formation of terrestrial planets \citep{lunine2011}.

\begin{figure*}
  \begin{center}
    \includegraphics[angle=270,width=16.0cm]{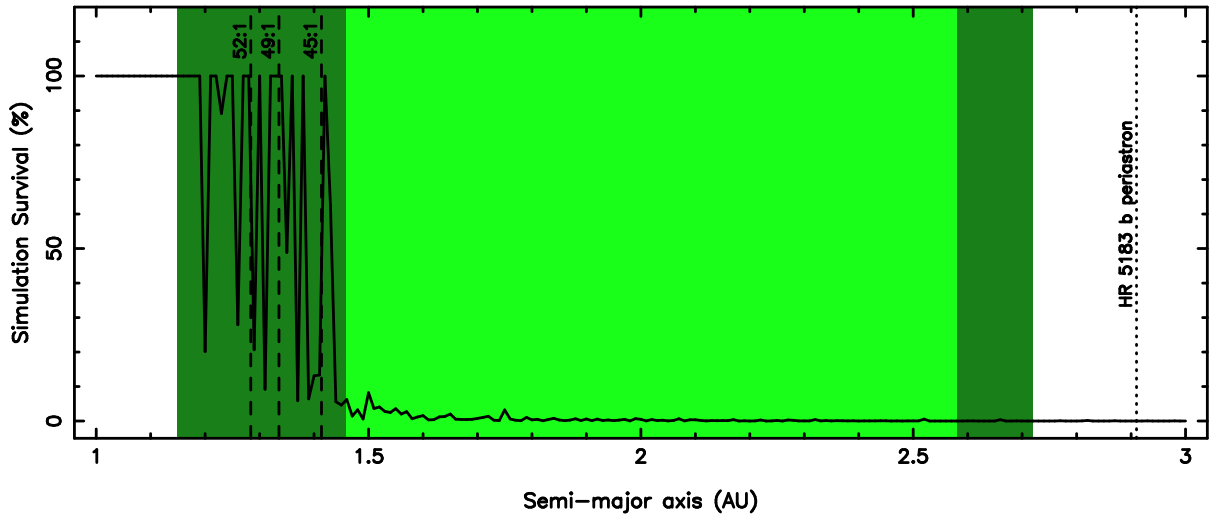}
  \end{center}
  \caption{Plot of the dynamical simulation results for an Earth-mass
    planet located between 1 and 3~AU from the host star, represented
    as the percentage survival of the simulation as a function of
    semi-major axis (solid line). As for Figure~\ref{systemfig}, the
    conservative HZ is shown in light green and the optimistic
    extension to the HZ is shown in dark green. The periastron passage
    of HR~5183b is indicated by the vertical dotted line.}
  \label{simfig}
\end{figure*}

To test for coplanar stable terrestrial planet locations within the
system, we conducted a suite of simulations that explore 200 evenly
spaced semi-major axes in the range 1.0--3.0~AU. This range was chosen
to fully encompass the extent of the optimistic HZ of the system (see
Section~\ref{system}). At each of the 200 semi-major axes, we placed
an Earth-mass planet at randomized starting positions (mean
anomalies). The dynamical simulations were then propagated in time
along with the known eccentric giant planet.

Our orbital stability simulations were undertaken using N-body
integrations with the Mercury Integrator Package \citep{chambers1999}.
The integrations utilized the hybrid symplectic/Bulirsch-Stoer
integrator with a Jacobi coordinate system, since that generally
provides more accurate results for multi-planet systems
\citep{wisdom1991,wisdom2006b} except in cases of close encounters
\citep{chambers1999}. Because relatively long orbital periods are
involved, we ran the simulations for $10^8$ years, commencing at the
present epoch and an orbital configuration output every 100 simulation
years. Based on the recommendations of \citet{duncan1998}, a time
resolution of 1.0~day was used to meet the minimum required resolution
of $1/20$ of the shortest orbital period within the system. The orbit
of the terrestrial planet is considered stable if it is able to retain
its orbital integrity for the duration of the simulation (i.e.,
neither ejected from the system nor lost to the potential well of the
host star).

The results of the dynamical simulations are summarized in
Figure~\ref{simfig}. At each of the test locations (semi-major axes)
for the terrestrial planet, the percentage of the simulations that
survived the full $10^8$~years are plotted. The solid line thus
indicates the stability of terrestrial planets as a function of
semi-major axis. As in Figure~\ref{systemfig}, the conservative HZ is
shown in light green and the optimistic extension to the HZ is shown
in dark green. The vertical dotted line at the right of the figure
indicates the periastron location of HR~5183b, and thus represents the
closest approach of the outer planet to the HZ. As can be seen from
the figure, the wrecking ball nature of the giant planet has a
devastating effect on the terrestrial planet stability regime,
rendering vast swathes of the HZ unstable. The bastion of hope for HZ
planets within the system lies within the inner optimistic HZ region,
where islands of stability reside. Some of these stability islands may
correspond to tenuous resonance locations, indicated by the vertical
dashed lines in Figure~\ref{simfig}. However, note that planets within
the inner optimistic HZ region may be Venus analogs rather than
temperate planets \citep{kane2014e}.

A further consideration with regards to stable terrestrial orbits is
that terrestrial planets can be significantly more massive than an
Earth-mass. The mass ratio of an Earth-mass planet to the mass of the
known planet (3.23~$M_J$) is $\sim$$10^{-3}$. Thus it is not expected
that a terrestrial planet would significantly influence the orbit of
the giant planet. To test the validity of these results for higher
mass planets, we repeated the simulations at several specific
star--planet separations for a five Earth-mass planet. These
simulation results were almost identical to those for the Earth-mass
planet simulations, confirming that the stability results presented
here apply to a broad range of terrestrial planet masses.

It is worth noting that the uncertainties for the orbital period and
semi-major axis of HR~5183b are relatively large. Here we have
explored the specific scenario surrounding the maximum likelihood
values of these parameters provided by \citet{blunt2019}. The planet
has recently passed through periastron passage where the greatest
constraints on the orbital solution can be made. Further data past
periastron passage over the coming years will allow for an improved
precision on the orbital period and subsequent dynamical implications.


\section{Predictions of Possible Additional Planets}
\label{predict}

The possibility of additional planets in the HR~5183 system must
satisfy the criteria of being both dynamically stable and beneath the
current data analysis detection thresholds. \citet{blunt2019} adopt a
planet injection and recovery technique to determine the sensitivity
of their data to additional planets. Their analysis shows that their
data are not sensitive to Earth-mass planets and their detection
threshold lies at $\sim$30~$M_\oplus$ in the range of
1--3~AU. Therefore, the possibility of terrestrial planets within the
stable orbit regimes described in Section~\ref{stability} remains
viable.

\begin{figure}
  \includegraphics[angle=270,width=8.5cm]{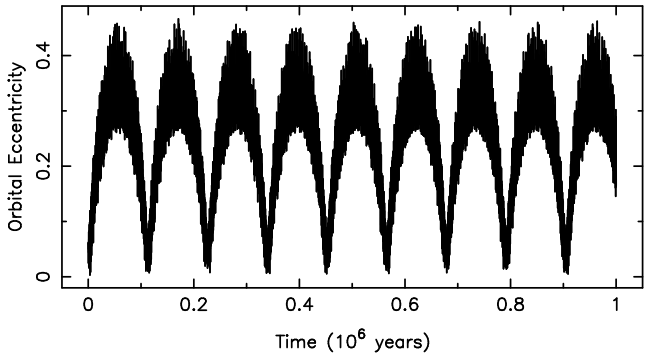}
  \caption{The eccentricity of a hypothetical terrestrial
    planet located at a semi-major axis of 1.42~AU as a function of
    time for $10^6$ simulation years. The plot shows variation of
    eccentricity in the range 0.0--0.5 as it exchanges angular
    momentum with the outer planet.}
  \label{eccfig}
\end{figure}

Here we consider an Earth-mass planet at the largest stable semi-major
axis; 1.42~AU from the host star. In order to further investigate the
orbital integrity at that location, we examined the time-dependent
eccentricity of the hypothetical planet, hereafter referred to as
planet c. The eccentricity of planet c over the first $10^6$~years of
the simulation is plotted in Figure~\ref{eccfig}. The figure shows
that the influence of the outer planet has a dramatic effect on the
planet c orbital eccentricity, causing it to vary in the range of
0.0--0.5 with a period of $\sim$$10^5$~years. Furthermore, the
high-frequency oscillations present in the eccentricity data match the
orbital period of the outer planet. The exchange of angular momentum
between two planets resulting in such eccentricity oscillations is
typical of systems where the two planets have diverse eccentricities
\citep{kane2014b}. However, in this case the angular momentum exchange
has a significant impact on the eccentricity of planet c whilst the
eccentricity of planet b remains largely unchanged.

\begin{figure}
  \includegraphics[angle=270,width=8.5cm]{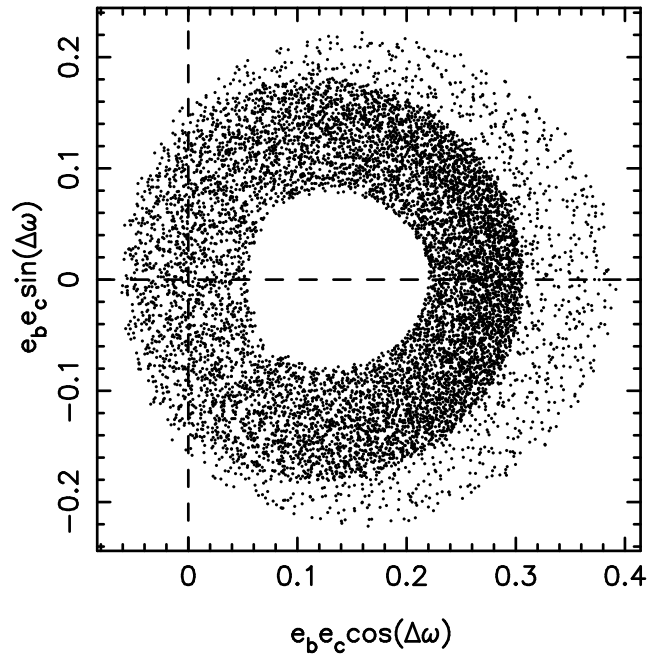}
  \caption{A polar plot of $e_b e_c$ vs $\Delta \omega$, representing
    the apsidal trajectory of the b and c planets, where ``b'' is the
    known giant planet and ``c'' is a hypothetical terrestrial planet.
    These data include the first $10^6$~years of the case where planet
    c is located at 1.42~AU from the host star. The figure shows that
    the apsidal modes are circulating during the dynamical evolution
    of the simulation, but lie close to the separatrix.}
  \label{epsilonfig}
\end{figure}

A further diagnostic of the eccentricity behavior combined with
overall long-term stability lies in the examination of the apsidal
mode trajectories. Apsidal motion is generally described as libration
or circulation, where the boundary between them is called a secular
separatrix \citep{barnes2006a,barnes2006c}. Shown in
Figure~\ref{epsilonfig} are the apsidal trajectories for planets b and
c, represented graphically in polar form. As in Figure~\ref{eccfig},
the data included are for the first $10^6$~years of the 1.42~AU
simulation. The data encompass the polar origin and so the planetary
system circulates. However, the distance of the apsidal trajectories
to the origin is small and so the system is subsequently close to the
separatrix boundary between libration and circulation. This proximity
to the separatrix explains the relatively high-frequency eccentricity
oscillations for planet c observed in Figure~\ref{eccfig}.

\begin{figure}
  \includegraphics[angle=270,width=8.5cm]{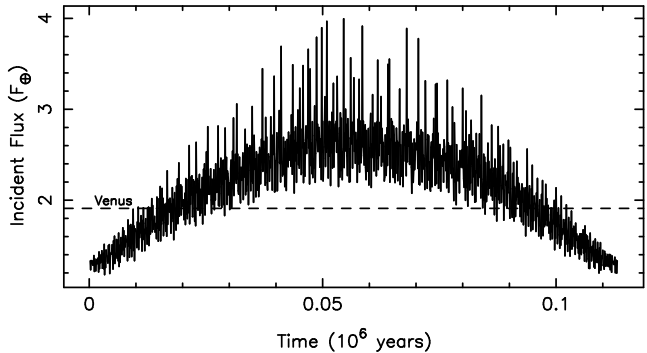}
  \caption{The variation in maximum incident flux received by the
    planet during one complete oscillation cycle of the eccentricity
    (see Figure~\ref{eccfig}) in units of the solar constant
    ($F_\oplus$). The dashed line indicates the average flux received
    by Venus from the Sun.}
  \label{fluxfig}
\end{figure}

As discussed in Section~\ref{intro}, giant planets within a system may
play a significant role in shaping the habitability of terrestrial
planets in those systems. As planet b moves through its eccentric
orbit, its perturbing influence will undoubtedly scatter material that
otherwise would have maintained long-term orbital stability. Many of
these perturbed objects will subsequently adopt orbits that make them
potential impactors on inner terrestrial planets
\citep{horner2008a,georgakarakos2018}. As shown in this section,
clearly a major effect of the wrecking ball nature of planet b on
HZ terrestrial planets are significant eccentricity variations. For
example, the variation in maximum incident flux received at the top of
the planetary atmosphere during a complete eccentricity oscillation
cycle is represented in Figure~\ref{fluxfig}. At the semi-major axis
of 1.42, the planet receives 1.17 times the solar constant
($F_\oplus$), but during periods of high eccentricity, the flux
received during periastron rises substantially above the amount of
flux received by Venus from the Sun, represented by the dashed line in
Figure~\ref{fluxfig}. Similarly, the planet would experience extended
periods of relatively low incident flux during the apastron. The
effects of orbital eccentricity on planetary climate with respect to
habitability have been investigated through the use of both simple and
complex climate simulations
\citep{williams2002,dressing2010,kane2012e,way2017a}. The change in
flux received by the planet during periods of high eccentricity would
result in eccentricity-driven seasonal effects rather than
obliquity-driven seasonal effects \citep{kane2017d}. However, the
thermal inertia of surface liquid water oceans can aid toward a
moderation of surface temperature variations and potentially mitigate
severe climate effects \citep{cowan2012c}. Even so, the tidal effects
caused by extreme eccentricity variations can in some cases lead to
runaway greenhouse scenarios resulting from the internal heating of
the planet \citep{barnes2013a}.

\begin{figure*}
  \begin{center}
    \includegraphics[angle=270,width=16.0cm]{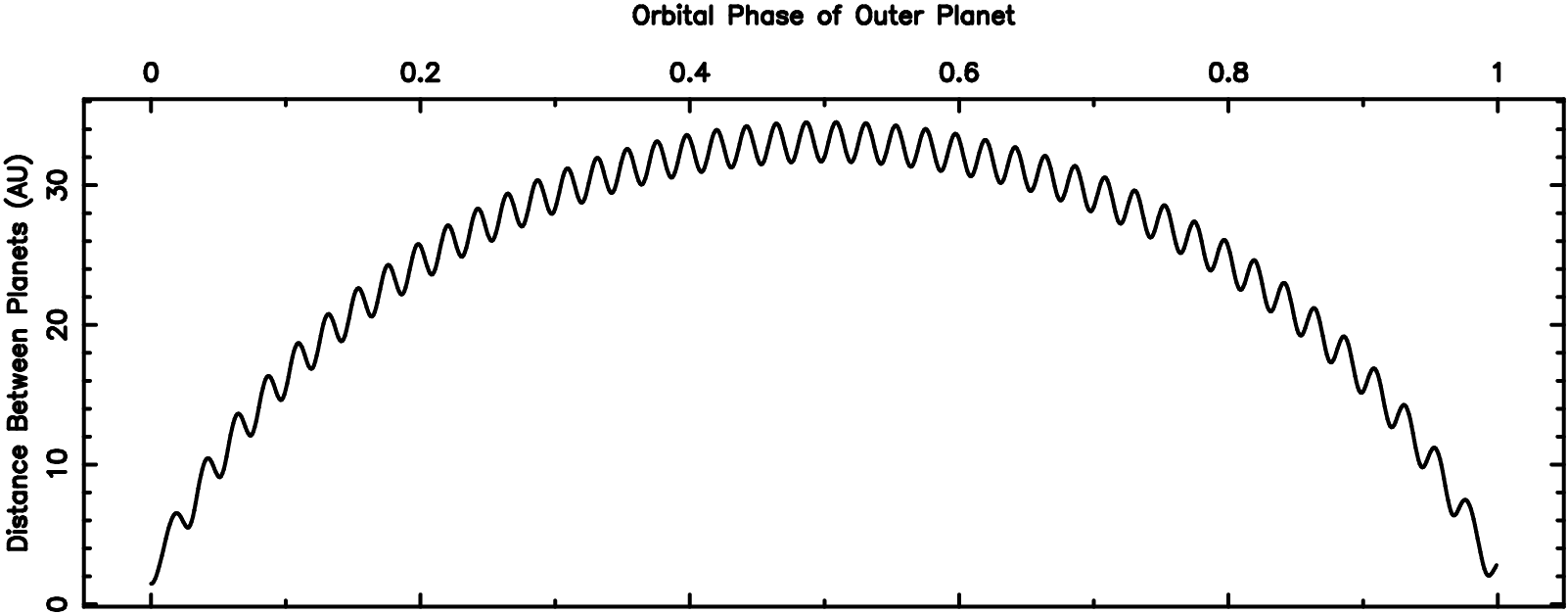} \\
    \includegraphics[angle=270,width=16.0cm]{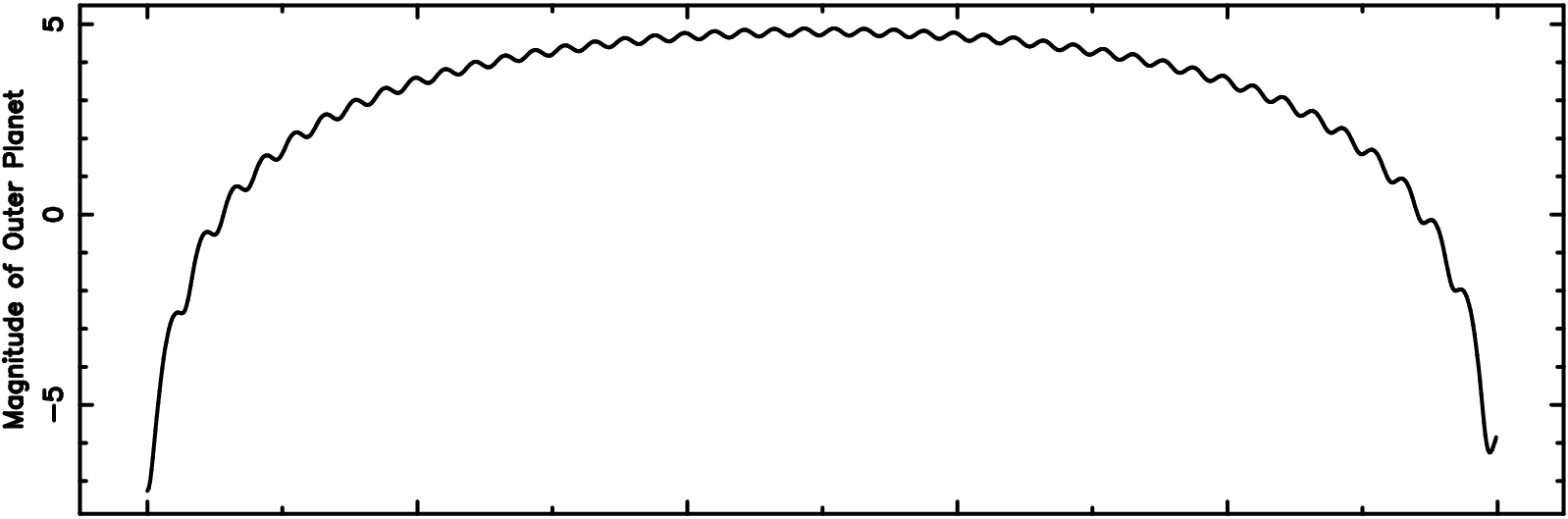} \\
    \includegraphics[angle=270,width=16.0cm]{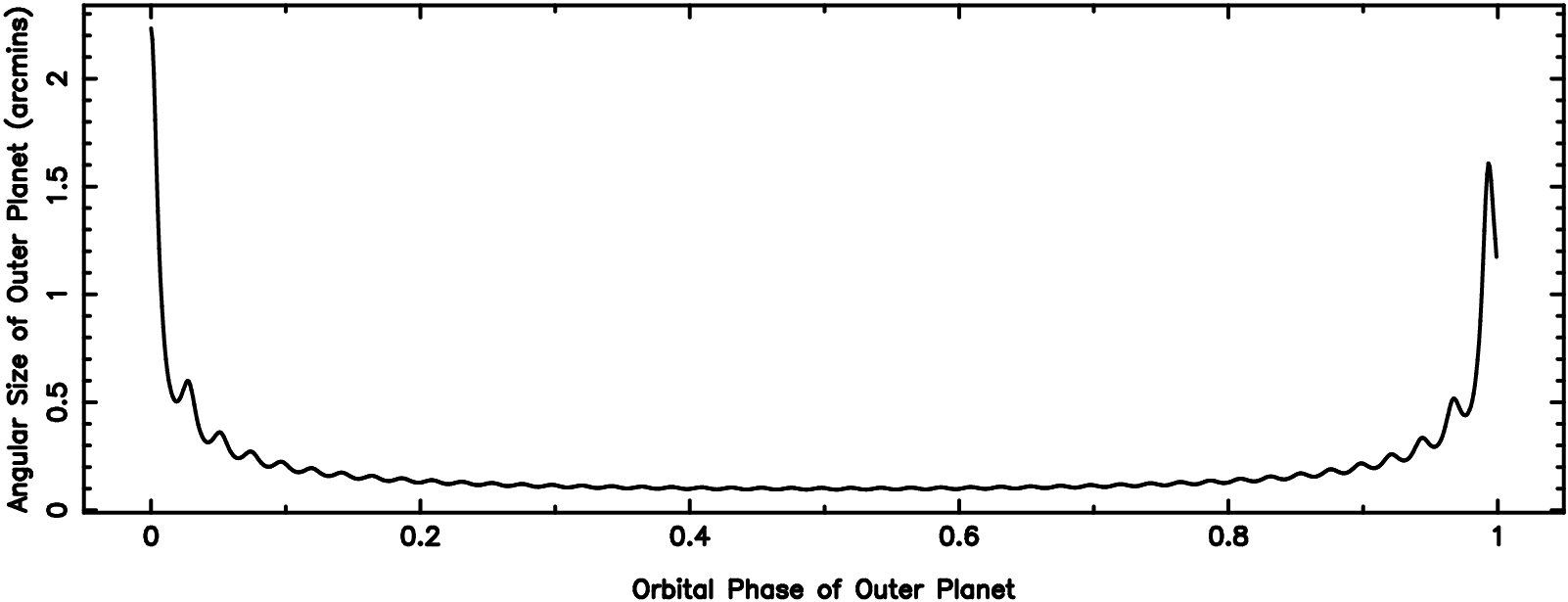}
  \end{center}
  \caption{Visibility of the known planet b from the perspective of a
    hypothetical terrestrial planet located at the farthermost stable
    orbit of 1.42~AU. All panels are shown as a function of the
    orbital phase of planet b where both of the planets are assumed to
    start (phase 0.0) at inferior conjunction. Top panel: Distance
    between both planets. Middle panel: Visible ($V$-band) magnitude
    of planet b. Bottom panel: Angular size of planet b.}
  \label{magfig}
\end{figure*}

A fascinating aspect of this system to explore is how planet b would
appear from the perspective of the hypothetical planet c over the
course of a complete orbit of planet b. If such a planet existed in
our solar system, the dramatic nature of the observable effects would
generally be considered a ``once in a lifetime'' event, similar to the
perihelion passage of Halley's comet. In Figure~\ref{magfig} we plot
the results of our calculations that explore the observable effects as
a function of the orbital phase of the outer planet. The top panel
shows the distance between the planets assuming that the planets both
begin at the minimum separation (inferior conjunction) of 1.47~AU. We
also assume that this interaction occurs during a period of time where
planet c occupies a circular orbit, as shown in
Figure~\ref{eccfig}. The farthest distance between the two planets
(superior conjunction) is 34.5~AU. The primary observable effect of
the change in distance will be the brightness of the outer planet,
depicted in the middle panel of Figure~\ref{magfig}, where we have
used a Jupiter geometric albedo of 0.538 for planet b. At closest
approach, the outer planet will have an apparent visual magnitude of
-7.3, 3 magnitudes brighter than Venus and as bright as the 1006
supernova (SN~1006). At its farthest distance, the outer planet has an
apparent visual magnitude of 4.9, still visible to the naked eye but
fainter than the open cluster M41. Another significant observable
change in the appearance of planet b during its orbit would be its
angular size, shown in the bottom panel of Figure~\ref{magfig}. At
closest approach, the angular diameter of the planet will be
2.2~arcminutes, compared with the 50~arcsecond size of Jupiter as seen
from the Earth during closest approach. This means that an average
person on planet c would be able to resolve the size of planet b
during the period of closest proximity.

\section{Conclusions}
\label{conclusions}

One of the primary surprises during the early discovery years of
exoplanets was the uncovering of giant planets on highly eccentric
orbits. The formation and subsequent orbital evolution of giant
planets can follow complex pathways when close encounters
significantly perturb the orbital stability of the
system. Fortunately, such orbital evolution does not always cause the
complete collapse of the system-wide orbital integrity, allowing other
planets to remain in stable orbits even in the presence of highly
eccentric giant planets. The discovery of particularly long-period
cases, such as HR~5183b, emphasizes the vast diversity of orbital
architectures that exist within an array of system formation and
outcome scenarios.

The importance of such systems from a planetary habitability
perspective arises from a thorough investigation of the dynamical
stability of terrestrial planetary orbits, such as the one presented
here. The careful analysis of the dynamical integrations demonstrates
that planets can survive within a narrow range of locations in the HZ
of such systems, even in the presence of a wrecking ball whose orbital
origin is likely a chaotic event involving vast exchanges of angular
momentum. However, the case of the HR~5183 system also shows that the
presence of an eccentric planet will often have a profound effect on
the Milankovitch cycles of the HZ terrestrial planetary orbits,
causing significant orbital oscillatory behavior. The implications for
the climate effects on such worlds may rule out temperate surface
conditions, although the stabilizing effects of surface liquid water
oceans can also potentially prevent a climate catastrophe. The
combination of dynamical simulations, more precise detection
techniques, and characterization of planetary atmospheres, will
eventually provide quantitative data to describe the full extent of
habitable environments in the presence of giant planets in extreme
orbits.


\section*{Acknowledgements}

The authors would like to thank Paul Dalba and Michelle Hill for
useful feedback on the manuscript. Thanks are also due to the referee,
Ravi Kopparapu, for his insightful comments. S.B. is supported by the
NSF Graduate Research Fellowship, grant No. DGE1745303. This research
has made use of the following archives: the Habitable Zone Gallery at
hzgallery.org and the NASA Exoplanet Archive, which is operated by the
California Institute of Technology, under contract with the National
Aeronautics and Space Administration under the Exoplanet Exploration
Program. The results reported herein benefited from collaborations
and/or information exchange within NASA's Nexus for Exoplanet System
Science (NExSS) research coordination network sponsored by NASA's
Science Mission Directorate.


\software{Mercury \citep{chambers1999}}




\end{document}